\documentclass[letter,scriptaddress, prl,showkeys,twocolumn]{revtex4}

	\usepackage[english]{babel}
	\usepackage[utf8]{inputenc}

	\usepackage{amsmath}%,amssymb} 
	\usepackage{makeidx}
	\usepackage{amsfonts}
\makeindex

\begin{document}

\title{Wave-particle duality from stochastic electrodynamics}

\author{Cesar Alberto Díaz Hernández}
\email{cadh@ciencias.unam.mx}
\affiliation{Facultad de ciencias, Universidad Nacional Autónoma de México, Apartado Postal 20-364, 01000 México, Distrito Federal, México}
\date{\today}

\begin{abstract}

The wave-particle duality is one of the most mysterious phenomena of the quantum theory, in this paper first it's studied the rise of the wave properties of matter from the theory of stochastic electrodynamics (SED), in which de Broglie's idea of a guiding wave that follows the particle and determine its trajectory is taken back. In the frame of the theory (SED) this wave is of electromagnetic character and is a traveling wave that propagates within the zero point field (ZPF) which according the SED is a classic electromagnetic field that permeates all universe. The wave rise from the interaction between the background field and the particle that moves with certain velocity. Then, I expose the results of numerical simulations in which the behavior of the ZPF in a two slit diffraction experiment is calculated and a series of macroscopic experiments that shows behaviors similar to the ones described by the SED. Both, simulations and experiments give us a better physical understanding of the guiding wave and of the wave properties of quantum particles

\end{abstract}


\maketitle

\section{Introduction}

The wave particle duality is one of the most important concepts within the quantum mechanics, the fact that particles such as electrons or neutrons exhibit a wave behavior can be seen in experiments such as the double slit experiment. Even that, the quantum theory describes perfectly this phenomena, it doesn't explain the reason why we can study one particle some times as a wave and others as a particle, hence, this duality is not very well understand, and some times misinterpreted
\\

In Luis de Broglie's papers where the matter wave mechanics is treated, it is not associated any physical meaning to the matter wave, nether an origin to such behavior \cite{de_broglie}. In later papers, he searched for a better understanding for the particle associated waves. He did not understand a particle as a wave; rather it is suggested the existence of a "pilot" wave that dictates the corpuscle trajectory. However, de Broglie didn't study the behavior of this wave and nether its interaction with the particle
\\

As basis for the SED it is assumed the existence of an electromagnetic the zero point field (ZPF) which is a classical background radiation field, i.e. that follows Maxwell's equations, with stochastic amplitude and it is assumed that its stochastic properties are the same for every inertial frame of reference \cite{E.Santos}. The stochastic electrodynamics (SED) theory suggest an interaction between a moving particle and the ZPF that give rise to an electromagnetic wave that rules the particle's trajectory, giving an explanation to the wave-particle duality as an emergent phenomena
\\

It is also assumed an intrinsic particle movement, known as "Zitterbewegung" witch is an oscillation for the particles with frequency $\omega_c$ close to Compton's frequency \cite{quantum_dice}

\begin{equation}
\omega_c  = \frac{m c^2}{\hbar}
\label{w_compton}
\end{equation}

This oscillation is such that the particle acquire an effective area and acts like it have an "internal watch". we notice that this oscillation interacts with the background field vibration's modes of frequency $\omega_c$

\section{De Broglie's wave}

Lets take a moving particle with a certain velocity $v$ immerse in the ZPF, as was mentioned before, the particle oscillates with frequency $\omega_c$, therefore it radiates, however, for the particle to be in equilibrium, this radiation most be absorbed by the field, i.e. the particle interacts with the field modes of frequency $\omega_c$ \cite{cetto1} 
\\

If we consider that the particle moves in the x direction, then, there will be two waves associated to the frequency $\omega_c$ each in one direction $-x$ and $x$. By the relativistic Doppler effect, this frequencies will be different. Lets be this frequencies $\omega_{+}$ and $\omega_{-}$, we can write

\begin{equation}
\omega_{\pm} = \omega_{z} \pm \omega_B
\label{omega_pm}
\end{equation}

Where $\omega_{z} = \gamma \omega_c$ is associated with the Lorentz's time contraction and $\omega_B = \gamma \beta \omega_c $ to the Doppler effect.
\\

Analogously we can define a wave number $k_{\pm}$ associated to $\omega_{\pm}$ as

\begin{equation}
k_{\pm} = \gamma k_z \pm k_B
\label{k_pm}
\end{equation}

with $k_B = \gamma \beta k_c$
\\

Thus, even if in the electron's frame of reference there is an stationary wave, in the laboratory frame of reference there is a superposition of waves with frequencies $\omega_{+}$ and $\omega_{-}$ that travels in opposite directions, lets $\phi_{-}$ and $\phi_{+}$ represents this waves, the resultant wave for this superposition will be given by

\begin{equation}
\phi(v) = \phi_{+} + \phi_{-} = 4 \cos{(\omega_z t - k_B x + \theta_{1}) \cos(\omega_B t- \gamma k_c x + \theta_{2})}
\label{phi}
\end{equation}

Where $\theta_1$ and $\theta_2$ are two random independent phases
\\

We notice that equation \eqref{phi} represents a traveling wave with space and time modulation. For the spatial case, we notice the traveling wave have wave number $k_B$, with a modulation for the wave length
 
\begin{equation}
\lambda_B = \frac{2 \pi}{k_B} = \frac{\lambda_c}{\gamma \beta} = \frac{m \lambda_c c}{p}
\label{mod_lambda}
\end{equation}

On the other hand

\begin{equation}
\lambda_c = \frac{2 \pi c}{\omega_c}
\label{lambda_c}
\end{equation} 

Thus, from equations \eqref{lambda_c} and \eqref{mod_lambda} we get

\begin{equation}
\lambda_B = 2 \pi \frac{m c^2}{\omega_c}\frac{1}{p}
\label{l_B}
\end{equation}

Finally, from equation \eqref{w_compton} it is obtain the following relation

\begin{equation}
\lambda_B = \frac{2 \pi \hbar}{p} = \frac{h}{mv}
\label{l_broglie}
\end{equation}

Equation \eqref{l_broglie} is the wave length equation for the matter waves suggested by Luis de Broglie. Thus, it is associated a traveling wave of electromagnetic character to the particle, we also notice that, the previous development does not depend on the particle's charge, which is why, the particle may be neutral. The ZPF polarize the neutral particle, creating an electromagnetic dipole, then, this particle interacts with the field as a particle with charge equal to the dipole charge

\section{Schrödinger's equation}

From SED and the "Zitterbewegung" it is possible to obtain Schrödinger's equation \cite{cavalleri}, however, most of this treatments requires an ensemble of particles. On the other hand, from the pilot wave from equation \eqref{phi} we can obtain time independent Schrödinger equation considering only one single particle
\\

Lets consider a wave equation for $\phi$ given by

\begin{equation}
\nabla^2  \phi - \frac{1}{u^2} \frac{\partial^2 \phi}{\partial t^2} =0
\label{eqonda}
\end{equation} 

Where $u$ is the wave $\phi$ phase velocity given by $u = c^2 / v$ with $v$ the particle's velocity
\\

from equation \eqref{phi}  we get

\begin{equation}
\frac{\partial^2 \phi}{\partial t^2} = -\omega_{z}^{2} \phi = - \frac{\gamma^2 m^{2} c^{4}}{\hbar^2} \phi
\end{equation} 

Thus, equation \eqref{eqonda} becomes

\begin{equation}
\nabla^2 \phi + \frac{v^2 \gamma^2 m^2}{\hbar^2} \phi = \nabla^2 \phi + \frac{p^2}{\hbar^2} \phi
\label{onda2}
\end{equation}

rewriting $p^2$  in terms of external potential $V$, we get

\begin{equation}
\nabla^2 \phi +\frac{1}{\hbar^2 c^2}[(\Sigma - V)^2 -m^2 c^4] \phi
\label{gordon}
\end{equation}

Where $\Sigma$ is the particle relativistic energy. we notice that equation \eqref{gordon} is Kein-Gordon's equation and in the non relativistic limit, with $E = \Sigma - m c^2$ we get

\begin{equation}
\nabla^2 \phi + \frac{2m}{\hbar}(E-V)\phi = 0
\label{schrodinger}
\end{equation}

This is time independent Schrödinger's equation for one particle. lets notices that, even if the equation make reference to the particle properties Such as its mass and energy, it was obtain from the equation of the wave that generates upon moving in the background field, showing that, even if they are different entities, there is a relationship between the pilot wave and the particle

\section{Physical meaning}

The SED bring us a new way to interpret matter waves, as shown before, de Brogle's wave is an electromagnetic wave that follows the moving particle. In this theory, the particle is always consider as a corpuscle, the wave-particle duality is not associated to it, however, its wave properties are given by the traveling wave that "guides" the particle within the background field, with this interpretation we obtain a more intuitive idea of the phenomena. Also, interpretations such as saying that "the particle interfere with it self" are avoided. Also, we notice that the pilot wave rises from the particle's velocity
\\

There are several interference and diffraction experiments with particles \cite{M.Surdin}, in general, this experiments consist in three phases, the fist one is the particle emission, the second one is the so called "interference screen" or "diffraction screen" and the third one is the detection screen
\\

For neutron interference, it is possible to use an arrangement such that neutron emission is one by one, i.e. one neutron is detected before the other is emitted, so that there isn't interference between two different neutrons. In this experiments we can preserve the wave properties of neutrons, and, we can conclude that a single neutron presents by its own, wave properties. Considering the stochastic theory given in this paper, that make sense, because a particle by its self already presents a pilot wave that rules its trajectory, in this sense, the particles don't interfere with their self and, neither they have "free will", simply, it's just the electromagnetic wave induced in the background field that presents interference (or is diffracted), this wave will be affected by such interference and, therefore, the particle as well, witch will have a bigger probability of being detected in the spots where the pilot wave's amplitude is bigger
\\

The mechanism by which, this wave rules the particle's behavior is not entirely clear, this is because of the difficulty of the mathematical theory, however, we can notice the works from J. Avenda\~no and L. de la pe\~na \cite{Avendano_1}, \cite{Avendano_2} in which the double slit experiment is studied and how its spatial conditions interact with the background field are considered. By numerical approximations it is possible to know the effects of the particles moving across the slits over the field, the resultant field is diffracted whit the intensity maximum distributed in a similar way that in a real life double slit experiment maximum, also, if the trajectories of the particles immerse in the ZPF are analyzed, it is notice that particles arrive to isolated spots on the screen (placed several wave lengths apart from the slit) being the regions with higher probability of a particle arriving the ones where the field (being diffracted) amplitudes are higher, this behavior is similar to Young's experiment for photons and for electrons. The results given by this simulations give us a better idea of how the ZPF give its wave properties to the particles and works as a more intuitive model of the wave-particle duality 
\\

Even that, to date, it is not possible for us to experimentally test the results of the SED given previously in this paper, there are similar macroscopic experiments, some of this experiments are the ones performed by Couder and Ford \cite{couder} in which a vertical oscillation is induced to the fluid, over it, little fluid drops bounce and move in directions perpendicular to the fluid vibrations, in this experiments it is notice that when the drops moves with a given velocity, a wave in the fluid surrounding the drop rises. This wave follows the droplet and guide its movement such as is described in the pilot wave theory for quantum particles. Also, by making the double slit experiment using the droplets it was observed a behavior similar to Young's experiment with electrons, each droplet pass trough some of the two slits, but, its associated wave passes trough both of them and it was diffracted. The final position of the droplets was always different and follows a statistic similar to a diffraction experiment. Although this experiments can't be used to verify the pilot wave theory for quantum particles, the behavior of fluid droplets bouncing in a medium with a given oscillation is similar to the particles interacting with the ZPF oscillation modes in witch are immersed

\section{Conclusion}

Under the frame of SED it was possible to give a more intuitive interpretation to the wave-particle duality, not as phenomena witch associate the particle and wave properties to the same object, instead, as two different and separated things the, wave and the particle related, but with each other. Thus, a particle never loses its corpuscular character and we could talk about a particle's trajectory, however, the particle's behavior within the ZPF is not easy to describe, given that is a stochastic field, is not possible to accurately predict the path that a particle will follow, however, with statistic treatments it can be said which will be the most probable position to find it given by the pilot wave intensity maximum, in that way, we can recover the already familiar results from quantum mechanics, including Schrodinger equation
\\

Even that there isn't a way to experimentally prove the pilot wave theory, the numeric simulations, the mentioned experiments in this paper and its similarities with the phenomena foretold by the SED lead us to thing that this description for quantum mechanics may be accurate, at least for the wave-particle duality phenomenon, but it would be necessary to perform more convincing experiments involving quantum particles so that we can certainly assure the existence of a pilot wave as described here.
\\

\bibliographystyle{unsrt}
\bibliography{biblio}

\end{document}